\documentclass[epj]{svjour}
\usepackage{graphics}
\usepackage{amssymb}
\begin{document}
%%%%%%%%%%%%%%%%%%%%%%%%%%%%%%%%%%%%%%%%%%%%%%%%%%%%%%%%%%%%%%%%%%%%%
\title{Self-healing slip pulses and the friction of gelatin gels}
\author{Tristan Baumberger \and Christiane Caroli \and Olivier Ronsin}
\offprints{ronsin@gps.jussieu.fr}
\institute{Groupe de Physique des Solides\thanks{associ\'e au Centre National de la Recherche Scientifique et aux Universit\'es Paris 6 et 7}, 2 place Jussieu, 75251 Paris, Cedex 05, France}
\date{Received: date / Revised version: date}

\abstract{
We present an extensive experimental study and scaling analysis of friction of gelatin gels on glass. At low driving velocities, sliding occurs via propagation of periodic self-healing slip pulses whose velocity is limited by collective diffusion of the gel network.
Healing can be attributed to a frictional instability occurring at the slip velocity $v = V_c$. For $v > V_c$, sliding is homogeneous and friction is ruled by the shear-thinning rheology of an interfacial layer of thickness of order the (nanometric) mesh size, containing a semi-dilute solution of polymer chain ends hanging from the network. Inspite of its high degree of confinement, the rheology of this system does not differ qualitatively from known bulk ones. The observed ageing of the static friction threshold reveals the slow increase of adhesive bonding between chain ends and glass. Such structural ageing is compatible with the existence of a velocity-weakening regime at velocities smaller than $V_c$, hence with the existence of the healing instability.
}
\maketitle
%%%%%%%%%%%%%%%%%%%%%%%%%%%%%%%%%%%%%%%%%%%%%%%%%%%%%%%%%%%%%%%%%%%%%
\section{Introduction}
\label{sec:intro}

When one shears a compliant medium forming a continuous macroscopic contact over a smooth hard substrate, sliding does not necessarily occur homogeneously. Schallamach\cite{Schallamach} made the first observation of an amazing inhomogeneous mode of sliding when studying the friction of rubber over smooth glass~: above a critical velocity, homogeneous sliding becomes unstable and the motion of rubber proceeds by propagation of periodic wrinkles that are detached from the substrate, the interface remaining stuck everywhere else. No measurable relative sliding occurs along the non detached interface, but each wrinkle is highly deformed (compressed) and provokes a finite relative displacement as it sweeps the interface.

Another inhomogeneous mode of sliding was proposed by Brune\cite{Brune} as a candidate for the relative motion of faults during an earthquake. It consists of a localized, slipping, non detached zone, bounded on both sides by stuck interface and propagating from the epicenter along the fault. This mode of rupture by propagation of a so-called self-healing slip pulse has regained attention from seismologists since the recent work of Heaton\cite{Heaton} who has shown that this description is compatible with observations for several large earthquakes.

A fundamental question is to identify the material surface and geometrical features which rule the existence and the dynamics of these modes of sliding. Despite of many experimental studies, it is still unclear which are the respective roles of the high deformability, viscoelasticity and adhesion hysteresis of elastomers in the formation of Schallamach waves\cite{Schallamach,Barquins}. As concerns the existence of slip pulses, it has been argued, on theoretical grounds\cite{Rice}, that a key point is the detailed nature of the friction law, and especially of the transition between the sticking and sliding regimes, i.e. of the pulse tip and resticking end. At this point, there is a clear need for systematic experimental investigation.

A few cases of observation of such self-healing slip pulses in laboratory experiments on soft systems have been reported~: polyurethane over araldite\cite{Villechaise}, and gelatin gels over glass\cite{Rubio}. We have recently revisited the latter system, and been able to provide the first detailed and quantitative analysis of the dynamics and internal structure of these pulses\cite{Ronsin}~: sliding is inhomogeneous up to a critical driving velocity $V_c$. In this regime, motion occurs via periodic stick-slip, each slip being associated with the propagation of a self-healing slip pulse nucleated at the trailing edge of the gel block. In contrast with the seismic case, the pulse velocity is strongly subsonic.

In the high driving velocity regime ($V > V_c$), steady homogeneous sliding sets in in the wake of an initial interfacial fracture which sweeps across the initially stuck contact. In this regime, the frictional stress is measured to be velocity-strengthening.

In the stick-slip regime, the measured slip velocity along the pulse decreases with distance behind the pulse tip, where it exhibits a crack-tip like behavior. When it reaches the value $V_c$ of the above mentioned critical velocity, it abruptly drops to zero, and the interface resticks. This suggests that the healing of the pulse is associated with a sliding instability occurring at velocity $V_c$, which could arise from a change of the velocity dependence of the friction stress from velocity-strengthening above $V_c$ to velocity-weakening below $V_c$.

It thus appears that a better understanding of the frictional properties of gelatin over glass is needed in order to get more insight into the mechanisms of this instability and thus of the formation of slip pulses. Osada {\it et al.}\cite{Gong} have studied in detail the normal load dependence of the friction stress for a number of gels on various substrates. However, systematic studies of the velocity dependence are still missing. In this article, we present an extensive study of gelatin$/$glass friction, extending our previous work\cite{Ronsin}, in which we investigate the effect on the dynamics of the gel characteristics, namely gelatin concentration and solvent viscosity.

On the basis of this study, we show that all the experimental results are compatible with the following picture of polymer gel-on-glass friction. The physical contact is formed by polymer blobs, hanging from the gel block, of height and lateral spacing comparable with the mesh size in the bulk. At rest, these chains get pinned to the glass through adhesive bonds, the adhesive strength increasing with time according to a slow dynamics. This ageing behavior manifests itself as a logarithmic increase of the threshold stress for fracture nucleation with time at rest. Upon sliding, bonds have only a finite lifetime and their number decreases with increasing sliding velocity $V$, since adhesive trapping is now limited by advection, an effect which has been modelled by several authors\cite{Schallamach2,Charitat,Gong2} who have shown it to be a source of velocity-weakening friction. At large enough velocities ($\gtrsim V_{\mathrm{c}}$) adhesive bonding becomes negligible and dynamic friction reduces to the viscous drag of a polymer/solvent layer of thickness the mesh size. This exhibits a power-law, shear-thinning behavior comparable with that of bulk polymer solutions\cite{Polymer}.

The paper is organized as follows~: section~\ref{sec:setup} describes the experimental setup and the preparation and characterization of our samples. In section~\ref{sec:frac}, we study systematically

(i) The dependence of the pulse tip velocity $V_{\rm tip}$ on the gel parameters. We give solid proof of the previously suggested\cite{Ronsin} result that $V_{\rm tip}$ is limited by energy dissipation via the diffusive mode of the poroelastic gel;
(ii) The interfacial ageing effect, namely the dependence of the fracture nucleation threshold on the sticking time;
(iii) The velocity and pressure dependence of the dynamic friction force in the homogeneous sliding regime.

The scaling analysis of these results, which substantiates the above scenario, is performed in section~\ref{sec:discuss}.

%%%%%%%%%%%%%%%%%%%%%%%%%%%%%%%%%%%%%%%%%%%%%%%%%%%%%%%%%%%%%%%%%%%%%

\section{Experimental set-up and gel characterization}
\label{sec:setup}

%%%%%%%%%%%%%%%%%%%%%%%%%%%%%%%%%%%%%%%%%%%%%%%%%%%%%%%%%%%%%%%%%%%%%

\subsection{Sample preparation}
\label{sec:sample}

\begin{figure}
\includegraphics{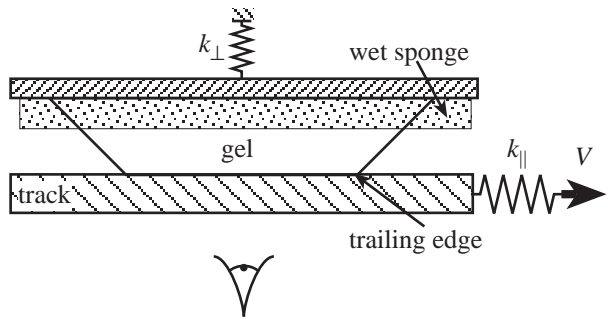}
\caption{Schematic setup of the experiment.}
\label{fig:setup}
\end{figure}

The system consists of a gel block sliding on a flat glass track as shown on Figure~\ref{fig:setup}. Gel samples are prepared from porcine skin gelatin\cite{Sigma} with concentration $c$ by weight ($c =$ 5, 8, 10 or 15\%) in mixtures of deionized water and a mass fraction $\phi$ of glycerol\cite{Prolabo} ($\phi =$ 0, 21 or 42\%). The pre-gel solution is heated under continuous stirring at a temperature of 90$^{\circ}$~C for 1 hour, then poured into a parallelipedic mold made of smooth glass with dimensions 5~cm$\,\times\,$1~cm$\,\times\,$1~cm. The mold covertop, later used as a holder for the gel, consists of a roughened micro-slide. The sample is kept at 5$^{\circ}$ C for 18 hours, then at room temperature (21 $\pm$ 1$^{\circ}$~C) for two hours. Once removed from the mold, the gel, which remains firmly attached to the rough slide, is cut into a trapezoidal shape (see Figure~\ref{fig:setup}) leading to a reduced free surface of area 3~cm$\,\times\,$1~cm. This shape, together with the roughness of the holding plate, ensures that, once contact has been established with the glass track, sliding does localize at this interface. The sample is surrounded with a piece of wet sponge which maintains a water saturated atmosphere, thus reducing the evaporation rate so that reproducible results are obtained when working with a given sample for more than one hour.

The track consists of a float glass plate. Its surface is cleaned using the following successive steps~: (i) degreasing with methanol, ethanol and acetone; (ii) sonicating in a 2\% solution of RBS detergent for 15 minutes at 50$^{\circ}$~C; (iii) rinsing and sonicating in ultrapure milli-Q water for 30 minutes; (iv) drying in oven.

%%%%%%%%%%%%%%%%%%%%%%%%%%%%%%%%%%%%%%%%%%%%%%%%%%%%%%%%%%%%%%%%%%%%%
%%%%%%%%%%%%%%%%%%%%%%%%%%%%%%%%%%%%%%%%%%%%%%%%%%%%%%%%%%%%%%%%%%%%%

\subsection{Setup}
\label{sec:meca}

The gel-holding slide is attached to one end of a double cantilever spring of stiffness $k_{\mathrm{\perp}}$ ($=  4.0\,10^3$ Nm$^{-1}$), the other end of which is position-controlled by a micrometric screw. The free surface of the gel block is set approximately parallel to the track, then brought into contact with it. Due to the strong gel$/$glass adhesion, as the two surfaces get close enough, the gel block jumps into contact. The adhesive contact is able to sustain negative loads, up to a finite tensile stress level at which detachment occurs via inward propagation of interfacial mode I fractures nucleating at the edges. Gel$/$track parallelism is finely adjusted by repeating this cycle until the fractures propagate symmetrically.

The track itself is attached to the end of a spring of stiffness $k_{\mathrm{\parallel}}$ ($= 1.8\,10^4$ Nm$^{-1}$), the other end of which is driven at a controlled velocity $V$ ranging between 1 $\mu$m$\,$s$^{-1}$ and 2 mm$\,$s$^{-1}$. Spring elongations are measured with the help of capacitive displacement transducers. Shear and normal stresses are thus measured with an accuracy of 10 Pa. The stiffnesses of both springs are more than ten times larger than of the gel block\footnote{Due to incompressibility on the timescale of the experiments, the compressive stiffness of the gel block is the undrained one, namely three times the shear stiffness.}. This entails that we impose the normal position of the holding rough slide and the velocity of the glass plate. The relative displacement between the contacting gel and glass surfaces is obtained by measuring optically, in the reference frame of the moving track, the position of tiny optical imperfections attached to the gel surface. We are thus able to measure the whole interfacial slip field with spatial and temporal resolutions of respectively 40 $\mu$m and 16 ms.

%%%%%%%%%%%%%%%%%%%%%%%%%%%%%%%%%%%%%%%%%%%%%%%%%%%%%%%%%%%%%%%%%%%%%

\subsection{Shear modulus}
\label{sec:modulus}

\begin{figure}
\includegraphics{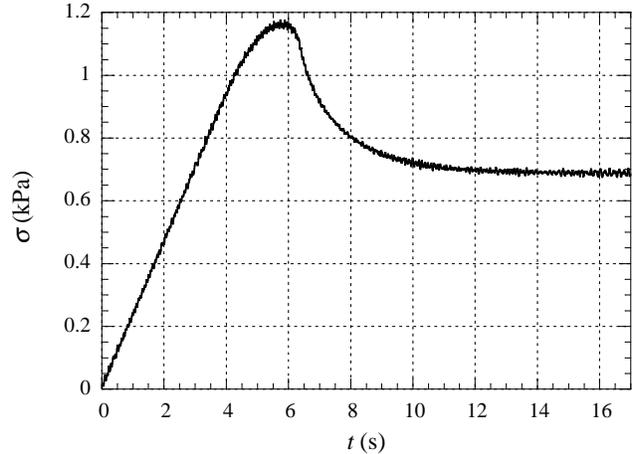}
\caption{Evolution of the shear stress when loading a gel block ($c = 5\%$, $\phi = 0$) from rest at the driving velocity $V = 500\ \mu$m.s$^{-1}$.}
\label{fig:static}
\end{figure}

The shear modulus $G$ of the gel is extracted from the stress response to shear displacement of the loading plate at constant velocity V. As long as the gel surface remains stuck to the glass track, the shear stress $\sigma$ increases linearly with time $t$ (Figure~\ref{fig:static}). Knowing the geometric characteristics of the gel block and assuming plane strain along the shear direction, one can easily relate the measured slope to the shear modulus~:
$$G = \frac{hl}{V}\frac{\ln (B/l)}{B-l}\frac{d\sigma}{dt}$$
with $l$ the length of the contacting surface, $B$ the length of the trapezoid base and $h$ the sample thickness. We thus measure shear moduli in the kPa range, which depend on the gel composition as shown in Table~1. Although gelatin forms physical gels which are known to strengthen with time by constantly forming new crosslinks\cite{Nijenhuis}, no evolution of $G$ is measured, up to experimental precision ($\simeq 5\%$), during the course of an experiment. This must certainly be assigned to the fact that bulk ageing is logarithmic only, while the duration of an experimental run ($\leq 1$ hour) is short on the scale of the previous history ($\simeq 20$ hours). Furthermore, in the velocity range used, no dependence of the modulus on the shear rate could be measured.

%%%%%%%%%%%%%%%%%%%%%%%%%%%%%%%%%%%%%%%%%%%%%%%%%%%%%%%%%%%%%%%%%%%%%
%%%%%%%%%%%%%%%%%%%%%%%%%%%%%%%%%%%%%%%%%%%%%%%%%%%%%%%%%%%%%%%%%%%%%

\subsection{Collective diffusion}
\label{sec:dls}

Gelatin gel is a two-component system made of a solid skeleton (polymer network) immersed in the solvent. Its mechanical properties are therefore those of a poroelastic medium. Namely, it exhibits, besides the ordinary acoustic modes of deformation, a third diffusive one, with dispersion relation $\omega = D_{\mathrm{coll.}}q^{2}$, which corresponds to relative motion of the two phases. The collective diffusion coefficient $D_{\mathrm{coll.}}$ characterizes the relaxation of concentration inhomogeneities of the gel network in the solvent. Tanaka {\sl et al.}~\cite{Tanaka} have shown that it could be measured from Dynamic Light Scattering (DLS) experiments.

\begin{figure}
\includegraphics{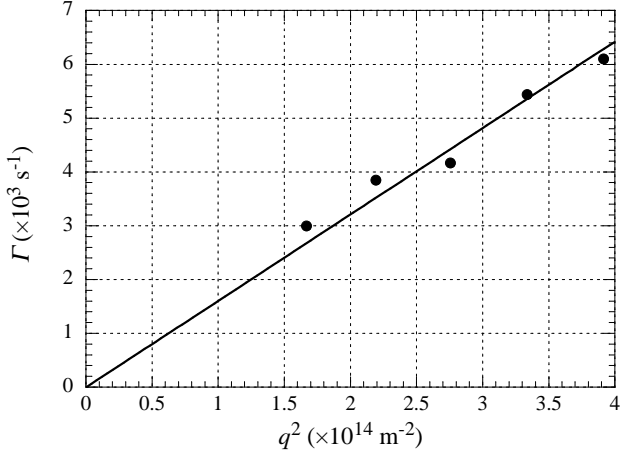}
\caption{Relaxation rate $\Gamma$ of the autocorrelation function of the intensity of light scattered \emph{vs} square of the scattering wavevector $q$ for a ($c = 5\%$, $\phi = 0$) sample. The linear relationship indicates diffusive relaxation.}
\label{fig:DLS}
\end{figure}

Gel samples are prepared in DLS cells in the same way as described above (section~\ref{sec:sample}). A Krypton laser ($\lambda_0 = 647$ nm) is focussed on the cell, maintained at the controlled temperature of 21$^{\circ}\pm 0.1$ C. The intensity of the light scattered at an angle $\theta$ between 60 and 120$^{\circ}$ from the incident beam is recorded with a photomultiplier. The auto-correlation function of the photocurrent is computed with a digital auto-correlator (BI-9000AT from Brookhaven Instrument Corporation). It decreases exponentially with a characteristic time $\tau$\cite{Pecora,Tanaka}. The relaxation rate $\Gamma = 1/2\tau$ is measured to increase linearily with $q^2$ as shown on Figure~\ref{fig:DLS}, where $q=(4\pi n/\lambda_0)\sin(\theta/2)$ is the scattering wavevector, $n$ the refraction index of the medium.

This is the behavior expected for the collective diffusive mode and the slope yields the collective diffusion coefficient $D_{\mathrm{coll.}}$. The values obtained for different gel compositions are reported in Table~1.

%%%%%%%%%%%%%%%%%%%%%%%%%%%%%%%%%%%%%%%%%%%%%%%%%%%%%%%%%%%%%%%%%%%%%
%%%%%%%%%%%%%%%%%%%%%%%%%%%%%%%%%%%%%%%%%%%%%%%%%%%%%%%%%%%%%%%%%%%%%

\subsection{Mesh size}
\label{sec:mesh}

\begin{figure}
\includegraphics{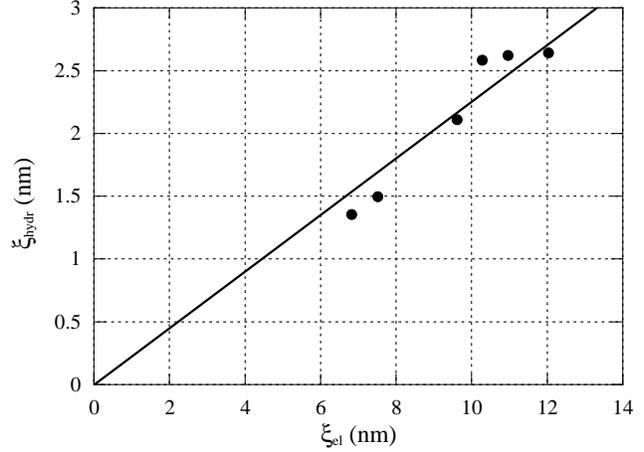}
\caption{Elastic \emph{vs.} hydrodynamic evaluations of the gel mesh size. The straight line shows the best linear fit.}
\label{fig:xi}
\end{figure}

The mesh size $\xi$ of the polymer network can then be estimated in two different ways~:
%%%%%%%%%%%%%%%%%%%%%%%%%%%%%%%%%%%%%%%%%%%%%%%%%%%%%%%%%%%%%%%%%%%%%
\begin{itemize}
\item An elastic estimate $\xi_{\mathrm{el}}$ can be obtained by making use of the scaling relation between the gel shear modulus and the distance $\xi$ between crosslinks\cite{PGG}~:
%%%%%%%%%%%%%%%%%%%%%%%%%%%%%%%%%%%%%%%%%%%%%%%%%%%%%%%%%%%%%%%%%%%%%
\begin {equation}
\label{eq:module}
G \sim \frac{k_{B}T}{\xi^{3}}
\end{equation}
%%%%%%%%%%%%%%%%%%%%%%%%%%%%%%%%%%%%%%%%%%%%%%%%%%%%%%%%%%%%%%%%%%%%%
from which we define
%%%%%%%%%%%%%%%%%%%%%%%%%%%%%%%%%%%%%%%%%%%%%%%%%%%%%%%%%%%%%%%%%%%%%
\begin{equation}
\label{eq:xiel}
\xi_{\mathrm{el}} = \left(\frac{k_{B}T}{G}\right)^{1/3}
\end{equation}
%%%%%%%%%%%%%%%%%%%%%%%%%%%%%%%%%%%%%%%%%%%%%%%%%%%%%%%%%%%%%%%%%%%%%
\item The continuum mechanics description of poroelastic gels leads to introducing a Darcy-like porosity $\kappa$, of order $\xi^{2}$, related with the collective diffusion coefficient $D_{coll}$ by\cite{Johnson}~:
%%%%%%%%%%%%%%%%%%%%%%%%%%%%%%%%%%%%%%%%%%%%%%%%%%%%%%%%%%%%%%%%%%%%%
\begin{equation}
\label{eq:darcy}
D_{\mathrm{coll.}} \simeq G\frac{\kappa}{\eta_{\mathrm{s}}}
\end{equation}
%%%%%%%%%%%%%%%%%%%%%%%%%%%%%%%%%%%%%%%%%%%%%%%%%%%%%%%%%%%%%%%%%%%%%
with $\eta_{\mathrm{s}}$ the solvent viscosity. From this, we define a hydrodynamic estimate of the mesh size~:
%%%%%%%%%%%%%%%%%%%%%%%%%%%%%%%%%%%%%%%%%%%%%%%%%%%%%%%%%%%%%%%%%%%%%
\begin{equation}
\label{eq:xihydr}
\xi_{\mathrm{hydr}} = \left(\frac{D_{\mathrm{coll.}}\eta_{\mathrm{s}}}{G}\right)^{1/2}
\end{equation}
%%%%%%%%%%%%%%%%%%%%%%%%%%%%%%%%%%%%%%%%%%%%%%%%%%%%%%%%%%%%%%%%%%%%%
\end{itemize}
The values of $\xi_{\mathrm{el}}$ and $\xi_{\mathrm{hydr}}$ for various gel concentrations and solvent compositions are listed in Table~1. As shown on Figure~\ref{fig:xi}, these two estimated values are indeed proportional, as expected. From now on, we will systematically use for $\xi$ the elastic estimate.

%%%%%%%%%%%%%%%%%%%%%%%%%%%%%%%%%%%%%%%%%%%%%%%%%%%%%%%%%%%%%%%%%%%%%
\begin{table*}
\label{tab:values}
\caption{Characteristics of different samples used.}
\begin{center}
\begin{tabular}{|c|c||c|c|c|c|c|}
\hline $c$ (wt.\%) & $\phi$ (wt.\%) & $G$ (kPa) & $D_{\mathrm{coll.}}$ ($\times 10^{11}$m$^2$/s) & $\xi_{\mathrm{el}}$ (nm) & $\eta_{\mathrm{s}}$ ($\times 10^{3}$Pa.s) & $V_{\mathrm{c}}$ ($\mu$m.s$^{-1}$) \\
\hline \hline 5 & 0 & 2.3 & 1.6 & 12 & 1.0 & 90\\
\hline 8 & 0 & 4.5 & 2.0 & 9.6 & 1.0 & 250\\
\hline 10 & 0 & 9.4 & 2.1 & 7.5 & 1.0 & 250\\
\hline 15 & 0 & 12.6 & 2.3 & 6.8 & 1.0 & 350\\
\hline 5 & 21 & 3.0 & 1.15 & 11.0 & 1.7 & 100\\
\hline 5 & 42 & 3.7 & 0.6 & 10.3 & 3.6 & 25\\
\hline
\end{tabular}
\end{center}
\end{table*}
%%%%%%%%%%%%%%%%%%%%%%%%%%%%%%%%%%%%%%%%%%%%%%%%%%%%%%%%%%%%%%%%%%%%%

%%%%%%%%%%%%%%%%%%%%%%%%%%%%%%%%%%%%%%%%%%%%%%%%%%%%%%%%%%%%%%%%%%%%%
%%%%%%%%%%%%%%%%%%%%%%%%%%%%%%%%%%%%%%%%%%%%%%%%%%%%%%%%%%%%%%%%%%%%%

\section{Experimental Results}
\label{sec:frac}
%%%%%%%%%%%%%%%%%%%%%%%%%%%%%%%%%%%%%%%%%%%%%%%%%%%%%%%%%%%%%%%%%%%%%
\subsection{Interfacial crack tip dynamics}
\label{sec:tip}
As already described in ref.~\cite{Ronsin}, when shearing the initially stuck gel block at constant velocity $V$, interfacial sliding always sets in via nucleation at the trailing edge of the gel of an interfacial shear crack, which propagates across the contact until it emerges at the leading edge. Unlike what occurs with the so-called Schallamach waves, no detachment (vertical opening of the contact) is observed optically behind the fracture tip. The fracture velocity was measured for each sample by following optically the tip position as a function of time\cite{Ronsin}. After nucleation, following an accelerating transient, too short for us to analyze it in detail, we observe that the fracture tip reaches a regime where it propagates at a constant velocity $V_{\mathrm{tip}}$, independent of the loading velocity in the range we have explored, where $V < V_{\mathrm{tip}}$. For all samples, it lies in the range of $1-20$ mm.s$^{-1}$. So, since shear wave velocities are of order m.s$^{-1}$, the interfacial fracture is strongly subsonic.

$V_{\mathrm{tip}}$ depends markedly on the gelatin concentration and solvent composition. We have suggested\cite{Ronsin} that the relevant limiting mechanism is energy dissipation at the pulse tip via collective diffusion mode, which becomes resonant which the depinning frequency when $V_{\mathrm{tip}}/\xi\sim D_{\mathrm{coll.}}/\xi^2$, i.e. for
%%%%%%%%%%%%%%%%%%%%%%%%%%%%%%%%%%%%%%%%%%%%%%%%%%%%%%%%%%%%%%%%%%%%%
\begin{equation}
\label{eq:Vtip}
V_{\mathrm{tip}} \sim \frac{D_{\mathrm{coll.}}}{\xi}
\end{equation}
%%%%%%%%%%%%%%%%%%%%%%%%%%%%%%%%%%%%%%%%%%%%%%%%%%%%%%%%%%%%%%%%%%%%%

\begin{figure}
\includegraphics{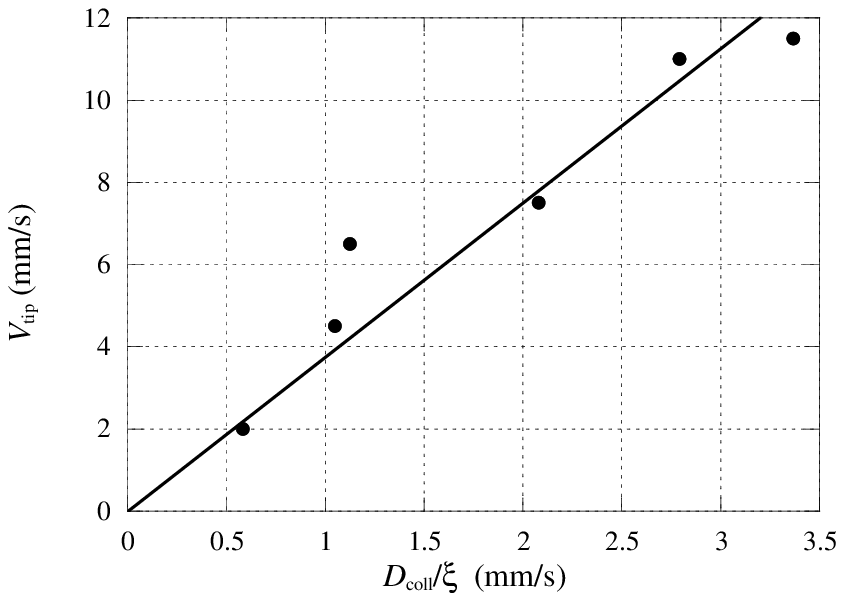}
\caption{Measured pulse tip velocity $V_{\mathrm{tip}}$ \emph{vs} $D_{\mathrm{coll.}}/\xi$ for different gel compositions.}
\label{fig:vfrac}
\end{figure}

By changing the gel composition, we were able to change both $D_{\mathrm{coll.}}$ and $\xi$ as shown on Table~1. Figure~\ref{fig:vfrac} shows the measured $V_{\mathrm{tip}}$ as a function of $D_{\mathrm{coll.}}/\xi$. We find that the proposed linear relationship is indeed verified, with a slope 2.8. This clearly justifies the scaling relation~(\ref{eq:Vtip}).

%%%%%%%%%%%%%%%%%%%%%%%%%%%%%%%%%%%%%%%%%%%%%%%%%%%%%%%%%%%%%%%%%%%%%

%%%%%%%%%%%%%%%%%%%%%%%%%%%%%%%%%%%%%%%%%%%%%%%%%%%%%%%%%%%%%%%%%%%%%
Tanaka \emph{et al.}\cite{Sekimoto} have shown that it is possible to drive a steady mode I fracture in a gel at velocities above $V_{D} \simeq D_{\mathrm{coll.}}/\xi$ which is therefore not an absolute dynamic limit. Whether our interfacial fractures might be forced beyond $V_{D}$, e.g. by loading the system at velocities larger than this value, remains a question open for future investigation. More generally, the qualitative nature of the argument that we have put forward to justify the observed velocity selection points to the interest of extending the theory of dynamic fracture to soft poroelastic materials such as gels, for which $V_{D}$ is strongly subsonic.

%%%%%%%%%%%%%%%%%%%%%%%%%%%%%%%%%%%%%%%%%%%%%%%%%%%%%%%%%%%%%%%%%%%%%
\subsection{Self healing slip pulses}
\label{sec:pulses}

\begin{figure}
\includegraphics{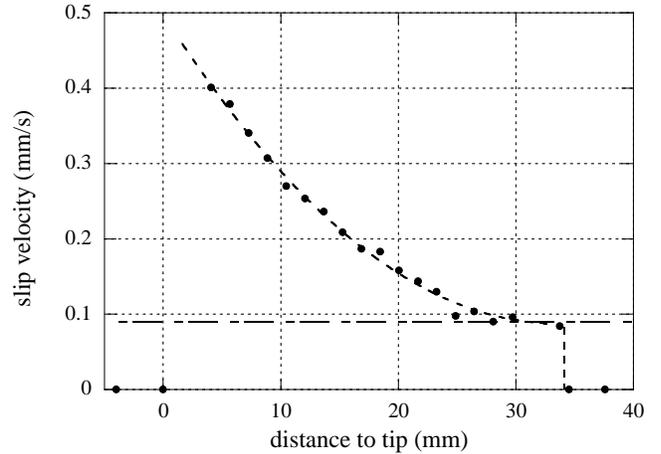}
\caption{Slip velocity \emph{vs} distance behind the crack tip for a ($c = 5\%$, $\phi = 0$) gel driven at $V = 10\ \mu$m$/$s. The dashed line is a guide for the eye. The dash-dotted line indicates the value $V_{\mathrm{c}} = 90\ \mu$m$/$s of the critical velocity.}
\label{fig:slipfield}
\end{figure}

While sliding is always initiated as described above, two different dynamical regimes are observed, depending on the value of the driving velocity~:
%%%%%%%%%%%%%%%%%%%%%%%%%%%%%%%%%%%%%%%%%%%%%%%%%%%%%%%%%%%%%%%%%%%%%
\begin{itemize}
\item for $V > V_{\mathrm{c}}$, the interfacial slip velocity decreases gradually with distance behind the tip, down to the asymptotic value $V$. After this transient, steady homogeneous sliding is established.
\item for $V < V_{\mathrm{c}}$, the local slip velocity close behind the tip exhibits the same type of variation. However, it does not decrease smoothly down to V, but, as seen on Figure~\ref{fig:slipfield}, it suddenly drops to zero, too fast for the corresponding transient to be optically resolved. Behind such a self-healing slip pulse, the interface resticks everywhere and gets elastically loaded again, until a new pulse nucleates. This results in a periodic stick-slip global dynamics, with stress drops typically on the order of $50 \%$ of the peak value.
\end{itemize}

%%%%%%%%%%%%%%%%%%%%%%%%%%%%%%%%%%%%%%%%%%%%%%%%%%%%%%%%%%%%%%%%%%%%%
%%%%%%%%%%%%%%%%%%%%%%%%%%%%%%%%%%%%%%%%%%%%%%%%%%%%%%%%%%%%%%%%%%%%%

It is important to note that resticking occurs when the \emph{local slip velocity} becomes equal to the critical value $V_{\mathrm{c}}$ of the \emph{driving velocity} which marks the transition of the global dynamics from stick-slip to steady sliding.

This behavior is also reflected in the fact that, the closer $V$ approaches $V_{\mathrm{c}}$ from below, the larger the pulse length and, accordingly, the duration of each slip phase and the corresponding slipped distance.

Note that the value of $V_{\mathrm{c}}$ strongly depends on the gel composition (Table~1). More precisely, it decreases markedly with increasing solvent viscosity, and increases with gelatin concentration.
% Valeurs de Vc dans le range 5---500$\mu$m/s dependance en\ldots

The fact that resticking occurs at a definite value of the local slip velocity, which does not depend on the driving one, but only on the gel poroelastic characteristics (see Table~1), strongly suggests that, in our system at least, self-healing is governed primarily by a frictional instability, while elastic interactions only play a secondary part. This is confirmed by the observation that, for $V$ close enough to $V_{\mathrm{c}}$, the pulse becomes longer than the sample itself, so that resticking occurs only after the pulse head has emerged out of the leading edge - i.e. when the tip elastic singularity has vanished.

Frictional sliding instabilities are usually associated with a regime where steady sliding exhibits a $V$-weakening regime. A direct check of such a behavior would require stiffening the system at will. In our case, this is impossible, since the most compliant element is the gel block itself. We must therefore rely upon the following indirect argument. It is well known that, in order to account for $V$-weakening, one must invoke the existence of some ``state variable'' describing the microscopic structure of the interface, the dynamics of which is non-instantaneous. This same dynamics usually also reflects into a time dependence of the interfacial strength, hence of the stress level for sliding initiation. This, in our case, is the fracture nucleation threshold.
%%%%%%%%%%%%%%%%%%%%%%%%%%%%%%%%%%%%%%%%%%%%%%%%%%%%%%%%%%%%%%%%%%%%%
\subsection{Ageing of the fracture threshold}
\label{sec:ageing}
\begin{figure}
\includegraphics{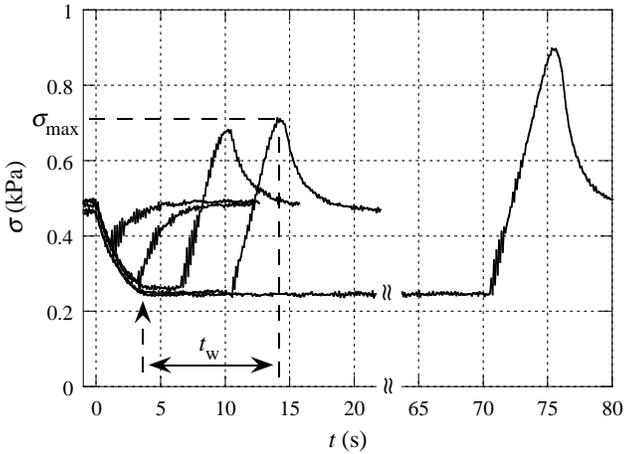}
\caption{A set of stop-and-go experiments for various loading histories ($c = 5\%$, $\phi = 0$). The arrow indicates the time at which the interface resticks as a whole. $\sigma_{\mathrm{max}}$ is defined as the level of the stress peak whenever it exists.}
\label{fig:stopgo}
\end{figure}

A convenient way to investigate a possible time dependence of the strength of the sticking interface is the so-called stop-and-go protocole (Figure~\ref{fig:stopgo}). Namely~: starting from steady sliding at some $V > V_{\mathrm{c}}$, the driving is stopped at time $t = 0$. The shear stress $\sigma$ then relaxes as the sliding velocity decreases until the latter reaches $V_{\mathrm{c}}$, at which point the interface resticks quasi-instantaneously, as reflected in the slope discontinuity of the $\sigma (t)$ curve (Figure~\ref{fig:stopgo}). This also appears consistent with our above assumption of a frictional healing instability. When loading at $V$ is resumed while the gel is still sliding, $\sigma$ increases smoothly back to its steady value. No transient peak nor overshoot is observed.

\begin{figure}
\includegraphics{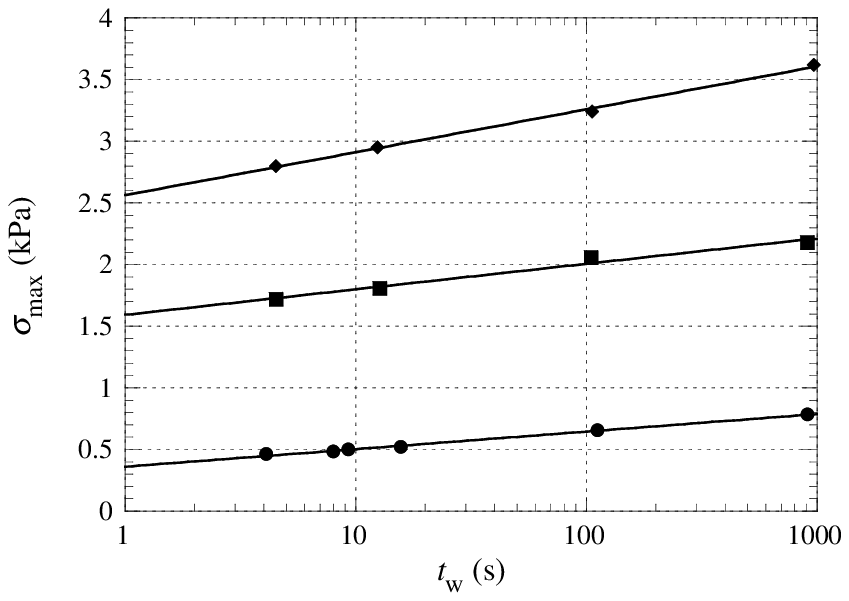}
\caption{Dependence of $\sigma_{\mathrm{max}}$ on waiting time at rest $t_{\mathrm{w}}$ for gelatin-water gels ($\phi = 0$) $\bullet$~: $c = 5\%$; $\blacksquare$~: $c = 8\%$; $\blacktriangle$~: $c = 10\%$. The loading velocity is $V = 300\ \mu$m/s. The thin lines are best linear fits.}
\label{fig:ageing}
\end{figure}

This contrasts with the response to reloading after resticking; in this case we always observe a transient peak, associated with the nucleation of a new fracture at the trailing edge. The height $\sigma_{\mathrm{max}}$ of this ``static peak'' increases with the waiting time $t_{\mathrm{w}}$ elapsed at rest, i.e. between resticking and slip initiation. More precisely, as shown on Figure~\ref{fig:ageing}, we find that $\sigma_{\mathrm{max}}$ increases logarithmically with $t_{\mathrm{w}}$, in the range 3 s $< t_{\mathrm{w}} \lesssim 10^{3}$ s, the lower limit being fixed by the loading duration.

We have studied static ageing for gels with concentrations 5, 8 and 10$\%$ of gelatin in water. As shown on Figure~\ref{fig:ageing}, the logarithmic slope of $\sigma_{\mathrm{max}}(t_{\mathrm{w}})$ is found to depend markedly on gelatin concentration. Note that we have not been able to perform the same systematic study with the other samples (15$\%$ gelatin + water, 5$\%$ gelatin + water/glycerol solvent). Indeed, in the latter cases, possibly due to the high level of $\sigma_{\mathrm{max}}$ already reached at the shortest waiting times, reloading after waiting at rest more than about 10 seconds usually resulted in fractures in the bulk.
%%%%%%%%%%%%%%%%%%%%%%%%%%%%%%%%%%%%%%%%%%%%%%%%%%%%%%%%%%%%%%%%%%%%%
%%%%%%%%%%%%%%%%%%%%%%%%%%%%%%%%%%%%%%%%%%%%%%%%%%%%%%%%%%%%%%%%%%%%%
%%%%%%%%%%%%%%%%%%%%%%%%%%%%%%%%%%%%%%%%%%%%%%%%%%%%%%%%%%%%%%%%%%%%%

\subsection{Frictional stress in homogeneous sliding}
\label{sec:friction}

\begin{figure}
\includegraphics{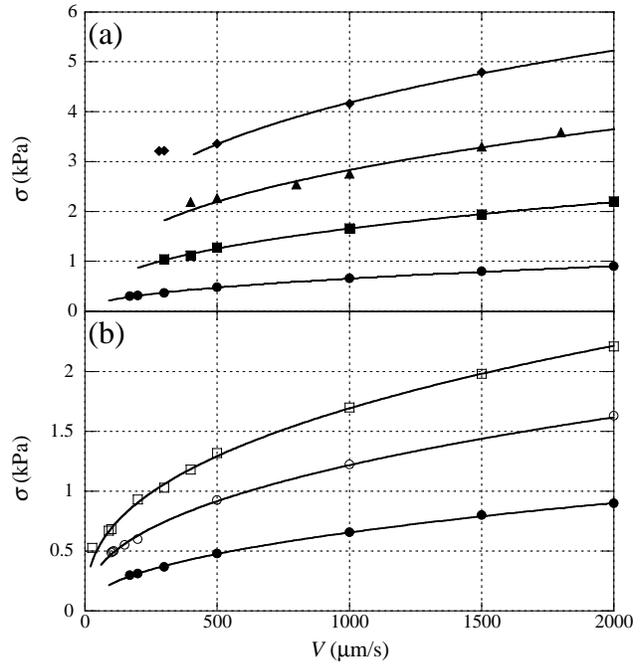}
\caption{Frictional stress $\sigma$ \emph{vs} driving velocity $V$ in the stationary sliding regime for different gel composition~: (a) gelatin in pure water ($\phi = 0$) $\bullet$~: $c = 5\%$; $\blacksquare$~: $c = 8\%$; $\blacktriangle$~: $c = 10\%$; $\blacklozenge$~: $c = 15\%$. (b) gelatin ($c = 5\%$) in water$/$glycerol mixture~: $\bullet$~: $\phi = 0$; $\circ$~: $\phi = 21\%$; $\square$~: $\phi = 42\%$. The thin lines are guides for the eye.}
\label{fig:rheo}
\end{figure}

For $V > V_{\mathrm{c}}$, after the initial transient, sliding becomes stationary and homogeneous, and is characterized by the velocity-dependent frictional stress $\sigma (V)$ which, as seen on Figure~\ref{fig:rheo}, depends markedly on the gel composition. In all cases, it clearly corresponds to a shear-thinning rheology for the sliding layer, that is, it grows with $V$ less than linearly.

%%%%%%%%%%%%%%%%%%%%%%%%%%%%%%%%%%%%%%%%%%%%%%%%%%%%%%%%%%%%%%%%%%%%%
%%%%%%%%%%%%%%%%%%%%%%%%%%%%%%%%%%%%%%%%%%%%%%%%%%%%%%%%%%%%%%%%%%%%%

\begin{figure}
\includegraphics{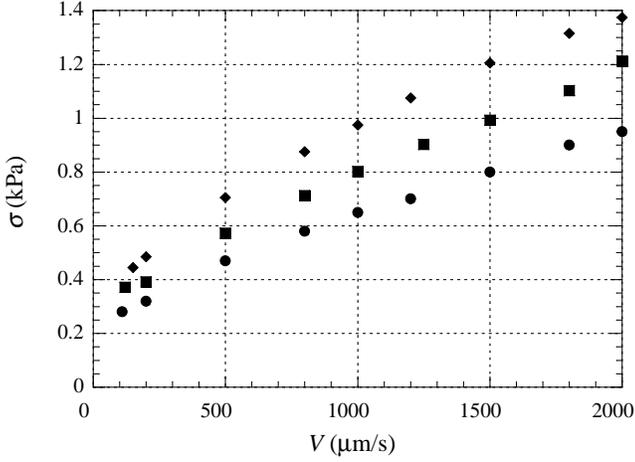}
\caption{Frictional shear stress $\sigma (V)$ ($c = 5\%$, $\phi = 0$) for different values of the normal stress $p$~: $\bullet$ $-0.15$ kPa; $\blacksquare$ 1.0 kPa; $\blacklozenge$ 1.65 kPa.}
\label{fig:norm}
\end{figure}

\begin{figure}
\includegraphics{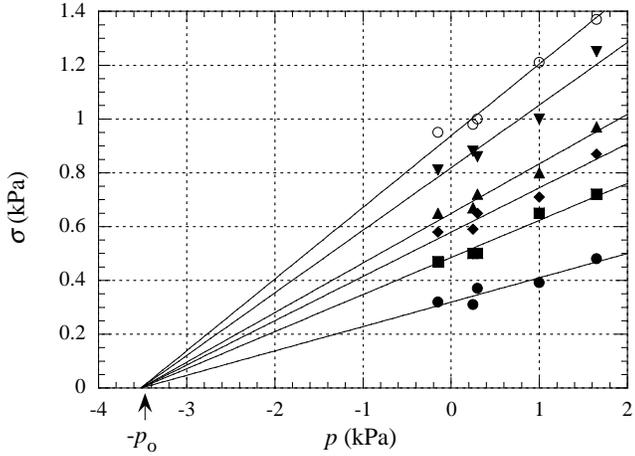}
\caption{Shear stress \emph{vs} normal stress ($c = 5\%$, $\phi = 0$). Each set of points corresponds to a given velocity~: $\bullet$ 200 $\mu$m$/$s; $\blacksquare$ 500 $\mu$m$/$s; $\blacklozenge$ 800 $\mu$m$/$s; $\blacktriangle$ 1000 $\mu$m$/$s; $\blacktriangledown$ 1500 $\mu$m$/$s; $\circ$ 2000 $\mu$m$/$s. The straight lines show the best linear fit for all sets with a common extrapolation normal stress $-p_0$.}
\label{fig:extrapol}
\end{figure}

We have also studied the dependence of $\sigma (V)$ on the applied normal stress $p$. It is illustrated, for a $5 \%$ gelatin~/ water gel, on Figure~\ref{fig:norm}. No measurable dependence of the critical velocity $V_{\mathrm{c}}$ on normal stress has been found. At any given velocity, $\sigma$ increases with $p$. This increase is quasi-linear (see Figure~\ref{fig:extrapol}) for normal stresses ranging from slightly negative (tensile) ones, accessible thanks to adhesion, up to compressive levels which may reach several tenths of the Young modulus. Moreover, for a given gel, the whole set of data can be fitted by the single expression~:
%%%%%%%%%%%%%%%%%%%%%%%%%%%%%%%%%%%%%%%%%%%%%%%%%%%%%%%%%%%%%%%%%%%%%
\begin{equation}
\label{eq:mu}
\sigma (V) = \mu(V) \left(p + p_{0}\right)
\end{equation}
%%%%%%%%%%%%%%%%%%%%%%%%%%%%%%%%%%%%%%%%%%%%%%%%%%%%%%%%%%%%%%%%%%%%%
where $p_{0}$ corresponds to the common extrapolation point shown on Figure~\ref{fig:extrapol}.

%%%%%%%%%%%%%%%%%%%%%%%%%%%%%%%%%%%%%%%%%%%%%%%%%%%%%%%%%%%%%%%%%%%%%
%%%%%%%%%%%%%%%%%%%%%%%%%%%%%%%%%%%%%%%%%%%%%%%%%%%%%%%%%%%%%%%%%%%%%
%%%%%%%%%%%%%%%%%%%%%%%%%%%%%%%%%%%%%%%%%%%%%%%%%%%%%%%%%%%%%%%%%%%%%
%%%%%%%%%%%%%%%%%%%%%%%%%%%%%%%%%%%%%%%%%%%%%%%%%%%%%%%%%%%%%%%%%%%%%

\begin{figure}
\includegraphics{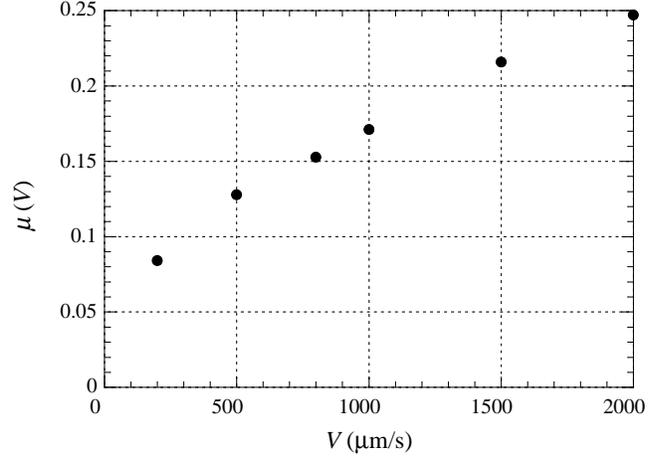}
\caption{Dynamic friction coefficient $\mu$ \emph{vs} sliding velocity $V$ ($c = 5\%$, $\phi = 0$).}
\label{fig:muV}
\end{figure}

The velocity dependence of the dynamic friction coefficient thus defined, $\mu(V)$, is displayed on Figure~\ref{fig:muV}.
%%%%%%%%%%%%%%%%%%%%%%%%%%%%%%%%%%%%%%%%%%%%%%%%%%%%%%%%%%%%%%%%%%%%%
%%%%%%%%%%%%%%%%%%%%%%%%%%%%%%%%%%%%%%%%%%%%%%%%%%%%%%%%%%%%%%%%%%%%%

%%%%%%%%%%%%%%%%%%%%%%%%%%%%%%%%%%%%%%%%%%%%%%%%%%%%%%%%%%%%%%%%%%%%%
\section{Scaling analysis and physical interpretation}
\label{sec:discuss}
In order to build a physical model of gelatin/glass friction, we will try to seek for possible scaling behaviors of the above data.

Let us first address the results concerning the fracture nucleation threshold. The fact that $\sigma_{\mathrm{max}}$ increases with time at rest immediately suggests that adhesion is due in part to the formation of bonds between gelatin blobs hanging from the network and the glass surface. Ageing can then naturally be assigned to the slow relaxation of the interfacial energy associated with reconfigurations of the confined and bonded polymer chains. Such dynamics, which involves a wide spectrum of relaxation times, is qualitatively consitent with the observed logarithmic behavior of interfacial ageing.

\begin{figure}
\includegraphics{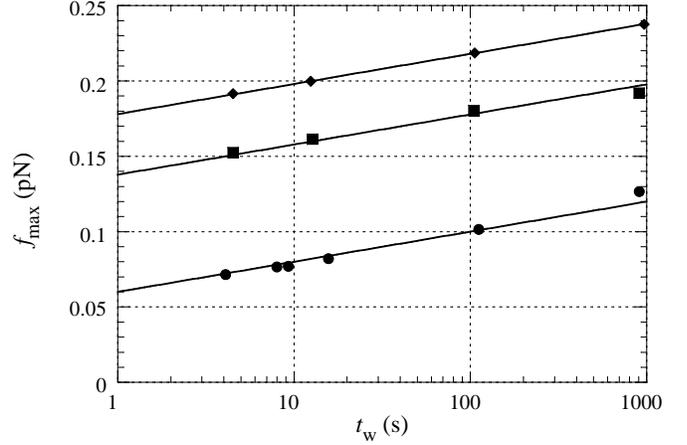}
\caption{Mean force per polymer blob $f_{\mathrm{max}} = \sigma_{\mathrm{max}}\xi^{2}$ \emph{vs} waiting time $t_{\mathrm{w}}$ for the same gels as in Figure~\ref{fig:ageing}. The straight lines are best logarithmic fits with a common slope.}
\label{fig:scaleage}
\end{figure}

It is reasonable to assume that the surface density of blobs candidate to bond formation scales as $\xi^{-2}$, and hence to try and plot the ageing data in terms, not of the threshold stress itself, but of the average force per blob $f_{\mathrm{max}} = \sigma_{\mathrm{max}}\xi^{2}$. The plot of Figure~\ref{fig:scaleage} indeed supports our interpretation, since it appears that the logarithmic slope of $f_{\mathrm{max}}(t_{\mathrm{w}})$ is independent of the gelatin concentration.

%%%%%%%%%%%%%%%%%%%%%%%%%%%%%%%%%%%%%%%%%%%%%%%%%%%%%%%%%%%%%%%%%%%%%
%%%%%%%%%%%%%%%%%%%%%%%%%%%%%%%%%%%%%%%%%%%%%%%%%%%%%%%%%%%%%%%%%%%%%

It is also seen that, the larger the gelatin concentration (the smaller $\xi$), the larger the upward shift of the $f_{\mathrm{max}}(t_{\mathrm{w}}) $ curves. We are unfortunately unable to analyze this effect in more detail. Indeed, it is clear that, in order for a physically meaningful offset to be defined, one should first of all scale, in the logarithmic ageing term, time $t$ by some characteristic $\tau(\xi)$ lying within the relaxation spectrum of the polymeric blobs. However, as long as we do not observe any saturation of the log variations in the time range we can access, we cannot identify a $\tau(\xi)$, and hence separate out an offset effect proper, if any.

Sliding necessarily results in limiting bond lifetime, which is then ruled by the balance between advection and rebonding dynamics --- a compound of elastic relaxation of blobs after bond breaking, diffusive exploration of the glass surface and duration of the bonding reactive process itself. Several authors \cite{Schallamach,Charitat,Gong2} have modelled this situation, and shown that it results in the existence of a velocity-weakening contribution to the steady sliding stress, which should vanish at high enough $V$'s, where advection becomes too fast for bonds to get formed. We believe that this is precisely the situation reached by our gelatin/glass systems already at velocities close above the critical value $V_{\mathrm{c}}$. In this $V$-strengthening regime, dynamic friction would therefore become entirely controlled by the viscous dissipation within the sheared interfacial layer.

\begin{figure}
\includegraphics{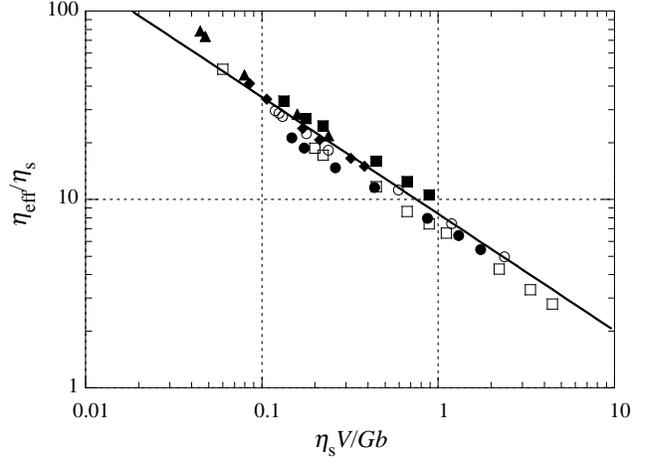}
\caption{Reduced effective viscosity $\eta_{\mathrm{eff}}/\eta_{\mathrm{s}}$ \emph{vs} Deborah number $\eta_{\mathrm{s}}V/G\,b$ for gels of different compositions. $b$ is abitrarily taken equal to 0.5 nm. The symbols are those of Figure~\ref{fig:rheo}.}
\label{fig:dynscale}
\end{figure}

In order to check this assumption, we define the effective viscosity $\eta_{\mathrm{eff}}$ of this layer, of thickness $\sim \xi$, thus sheared at the rate $\dot{\gamma} = V/\xi$, as
%%%%%%%%%%%%%%%%%%%%%%%%%%%%%%%%%%%%%%%%%%%%%%%%%%%%%%%%%%%%%%%%%%%%%
\begin{equation}
\label{eq:visc}
\eta_{\mathrm{eff}} = \frac{\sigma\xi}{V}
\end{equation}
%%%%%%%%%%%%%%%%%%%%%%%%%%%%%%%%%%%%%%%%%%%%%%%%%%%%%%%%%%%%%%%%%%%%%
We then plot all our dynamical data as $\eta_{\mathrm{eff}}/\eta_{\mathrm{s}}$ versus the Deborah number for a solution of gaussian blobs of size $\xi$ in a solvent of viscosity $\eta_{\mathrm{s}}$
%%%%%%%%%%%%%%%%%%%%%%%%%%%%%%%%%%%%%%%%%%%%%%%%%%%%%%%%%%%%%%%%%%%%%
$$\mathcal{D} = \dot{\gamma} \tau_{R}$$
%%%%%%%%%%%%%%%%%%%%%%%%%%%%%%%%%%%%%%%%%%%%%%%%%%%%%%%%%%%%%%%%%%%%%
with $\tau_{R}$ the Rouse time of a chain of $N$ monomers of size $b$~:
%%%%%%%%%%%%%%%%%%%%%%%%%%%%%%%%%%%%%%%%%%%%%%%%%%%%%%%%%%%%%%%%%%%%%
$$\tau_{R} = \frac{2 \eta_{\mathrm{s}}\xi^{4}}{\pi b\,k_{B}T}$$
%%%%%%%%%%%%%%%%%%%%%%%%%%%%%%%%%%%%%%%%%%%%%%%%%%%%%%%%%%%%%%%%%%%%%
so that
%%%%%%%%%%%%%%%%%%%%%%%%%%%%%%%%%%%%%%%%%%%%%%%%%%%%%%%%%%%%%%%%%%%%%
$$\mathcal{D} \sim \frac{\eta_{\mathrm{s}}V}{G\,b}$$
%%%%%%%%%%%%%%%%%%%%%%%%%%%%%%%%%%%%%%%%%%%%%%%%%%%%%%%%%%%%%%%%%%%%%
It is seen on Figure~\ref{fig:dynscale} that all data corresponding to different gelatin concentrations and glycerol fractions then collapse onto a single curve, well fitted by a power law, namely:
%%%%%%%%%%%%%%%%%%%%%%%%%%%%%%%%%%%%%%%%%%%%%%%%%%%%%%%%%%%%%%%%%%%%%
\begin{equation}
\label{eq:powerlaw}
\frac{\eta_{\mathrm{eff}}}{\eta_{\mathrm{s}}} \sim \mathcal{D}^{-\alpha}
\end{equation}
%%%%%%%%%%%%%%%%%%%%%%%%%%%%%%%%%%%%%%%%%%%%%%%%%%%%%%%%%%%%%%%%%%%%%
with the exponent $\alpha = 0.6 \pm 0.07$.

The existence of such scaling behavior, together with the fact that power-law shear thinning rheologies with comparable exponents are common for bulk polymer solutions, brings strong support to our above assumption.

%%%%%%%%%%%%%%%%%%%%%%%%%%%%%%%%%%%%%%%%%%%%%%%%%%%%%%%%%%%%%%%%%%%%%
%%%%%%%%%%%%%%%%%%%%%%%%%%%%%%%%%%%%%%%%%%%%%%%%%%%%%%%%%%%%%%%%%%%%%

\begin{figure}
\includegraphics{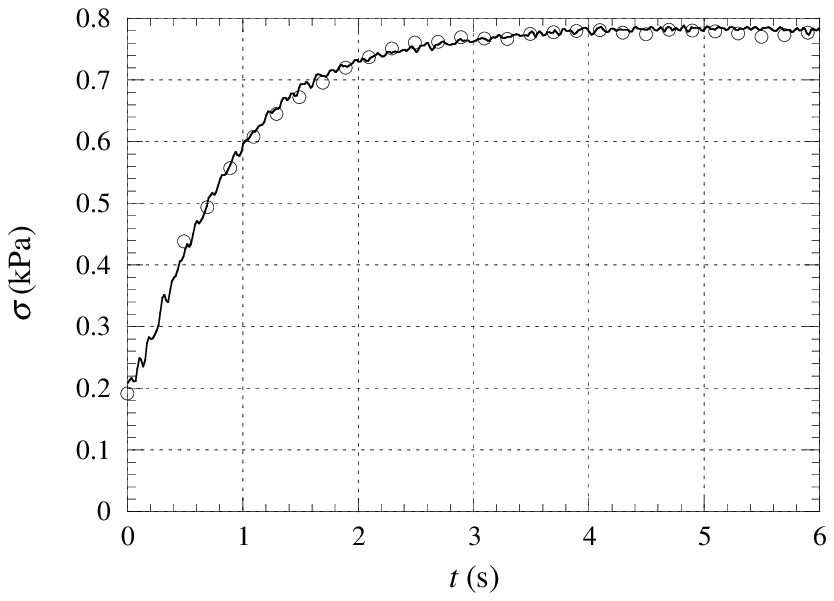}
\caption{Full curve : measured shear stress $\sigma$ \emph{vs} time following a jump at $t = 0$ of the driving velocity $V$ from 100 to 1800 $\mu$m$/$s. Dots : values of the stationary sliding stress calculated at the corresponding values of the instantaneous slip velocity according to the fit of Figure~\ref{fig:dynscale}.}
\label{fig:instant}
\end{figure}

Moreover, note that within this picture, the duration of friction transients following driving velocity jumps should be controlled by the Rouse times, typically of order $10^{-4}-10^{-5}$ s (taking $b$ on the order of a few Angstr\"oms), much shorter than both our time resolution and the response time of our gel blocks. We therefore expect that the frictional stress should appear, on experimental time scales, as adjusting instantaneously to sliding velocity variations. This we have checked by comparing the transient stress variations $\sigma(t)$ following driving velocity jumps with values of the \emph{stationary} stress $\sigma[V = \dot{x}(t)]$ corresponding to the measured instantaneous sliding velocity $\dot{x}(t)$. The agreement is excellent, as illustrated on Figure~\ref{fig:instant}.

%%%%%%%%%%%%%%%%%%%%%%%%%%%%%%%%%%%%%%%%%%%%%%%%%%%%%%%%%%%%%%%%%%%%%
%%%%%%%%%%%%%%%%%%%%%%%%%%%%%%%%%%%%%%%%%%%%%%%%%%%%%%%%%%%%%%%%%%%%%

Finally, let us turn to the dependence of the frictional stress on the normal load. In the frame of the above description, it is reasonable to consider that this dependence arises essentially from the elastic variations of the interfacial layer thickness under uniaxial compression. The variation $\delta\xi$ of the layer thickness under the normal compressive stress $p$ reads:
%%%%%%%%%%%%%%%%%%%%%%%%%%%%%%%%%%%%%%%%%%%%%%%%%%%%%%%%%%%%%%%%%%%%%
$$\frac{\delta\xi}{\xi} = -\frac{p}{E}$$
%%%%%%%%%%%%%%%%%%%%%%%%%%%%%%%%%%%%%%%%%%%%%%%%%%%%%%%%%%%%%%%%%%%%%
with $\xi$ the thickness under zero applied normal stress and $E = 3G$ the Young modulus.

\begin{figure}
\includegraphics{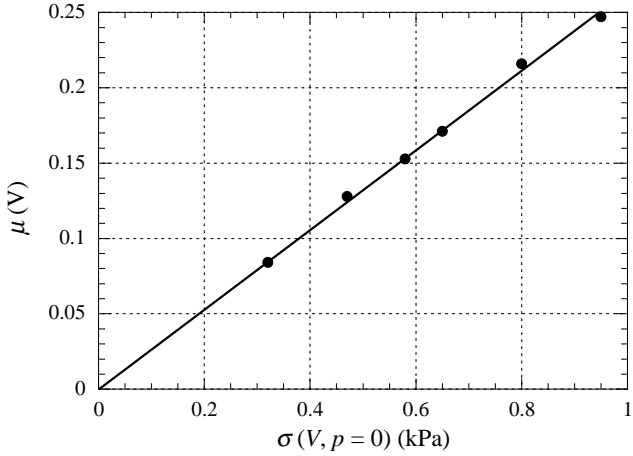}
\caption{Dynamic friction coefficient $\mu (V)$ \emph{vs} shear stress under zero normal load $\sigma (V, p = 0)$. Each point corresponds to a different velocity $V$.}
\label{fig:mucolle}
\end{figure}

As long as $\delta\xi/\xi \ll 1$, and making use of the power law rheology (eq.~(\ref{eq:powerlaw})), the relative frictional stress variation reads:
%%%%%%%%%%%%%%%%%%%%%%%%%%%%%%%%%%%%%%%%%%%%%%%%%%%%%%%%%%%%%%%%%%%%%
$$\frac{\delta\sigma}{\sigma} = -(3\alpha +1)\frac{\delta\xi}{\xi_{0}}$$
%%%%%%%%%%%%%%%%%%%%%%%%%%%%%%%%%%%%%%%%%%%%%%%%%%%%%%%%%%%%%%%%%%%%%
whence:
%%%%%%%%%%%%%%%%%%%%%%%%%%%%%%%%%%%%%%%%%%%%%%%%%%%%%%%%%%%%%%%%%%%%%
$$\sigma(V,p) = \sigma(V,p=0)\left\lbrack 1+(1+3\alpha)\frac{p}{3G}\right\rbrack$$
%%%%%%%%%%%%%%%%%%%%%%%%%%%%%%%%%%%%%%%%%%%%%%%%%%%%%%%%%%%%%%%%%%%%%
This reduces to the form of Equation~(\ref{eq:mu}) provided that one sets~:
%%%%%%%%%%%%%%%%%%%%%%%%%%%%%%%%%%%%%%%%%%%%%%%%%%%%%%%%%%%%%%%%%%%%%
\begin{equation}
\label{eq:p0}
p_{0} = \frac{3G}{1+3\alpha} \approx G
\end{equation}
%%%%%%%%%%%%%%%%%%%%%%%%%%%%%%%%%%%%%%%%%%%%%%%%%%%%%%%%%%%%%%%%%%%%%
and
%%%%%%%%%%%%%%%%%%%%%%%%%%%%%%%%%%%%%%%%%%%%%%%%%%%%%%%%%%%%%%%%%%%%%
\begin{equation}
\label{eq:musigma}
\mu(V) = \sigma(V,p=0)/p_{0}
\end{equation}
%%%%%%%%%%%%%%%%%%%%%%%%%%%%%%%%%%%%%%%%%%%%%%%%%%%%%%%%%%%%%%%%%%%%%
In the case displayed on Figure~\ref{fig:extrapol}, $p_{0} \approx 3.4$ kPa while $3G/(1+3\alpha) \approx 3 \pm 0.2 $ kPa. As seen on Figure~\ref{fig:mucolle}, for the same sample, $\mu (V)$ varies linearly with $\sigma (V, p = 0)$, as predicted by equation~(\ref{eq:musigma}). The experimental slope is $0.26$ kPa$^{-1}$, to be compared with $p_{0}^{-1} \approx 0.29$. That is, agreement is in this case very satisfactory. The other gelatin/water samples for which we have studied the effect of normal stress all had larger gelatin concentration, hence larger values of $G$. This makes the determination of the extrapolation stress $-p_{0}$ all the less accurate, and all we can afford to state is that $p_{0}$ does increase with $G$.\newline

%%%%%%%%%%%%%%%%%%%%%%%%%%%%%%%%%%%%%%%%%%%%%%%%%%%%%%%%%%%%%%%%%%%%%
%%%%%%%%%%%%%%%%%%%%%%%%%%%%%%%%%%%%%%%%%%%%%%%%%%%%%%%%%%%%%%%%%%%%%

In summary, we have been able to interpret gelatin-glass frictional dissipation as taking place within a layer of thickness of order the mesh size $\xi$, made of a semi-dilute solution of polymer chains attached to the network. At rest, these chains adhere to the glass substrate, and the interfacial strength, as measured by its shear fracture threshold, ages logarithmically. The existence of a low-velocity dynamics consisting of periodic slip pulses which heal at a critical value of the local slip velocity is compatible with a $V$-weakening frictional regime, which we associate with the decrease with shear rate of the adhesive bonding strength. Above $V_{\mathrm{c}}$, adhesion becomes negligible, and frictional dissipation is ruled by the shear-thinning rheology of the layer. It is remarkable that, inspite of its nanometric thickness, this rheology does not differ qualitatively from what is known for bulk polymer solutions.

We believe that the results of this study, while limited to soft poroelastic gels, point towards the importance of understanding the detailed behavior of frictional laws for unraveling of the question of the existence and dynamical characteristics of slip pulses.

\begin{acknowledgement}
We thank B. Velick\'y for his help in the analysis of the data.
\end{acknowledgement}

%%%%%%%%%%%%%%%%%%%%%%%%%%%%%%%%%%%%%%%%%%%%%%%%%%%%%%%%%%%%%%%%%%%%%

%%%%%%%%%%%%%%%%%%%%%%%%%%%%%%%%%%%%%%%%%%%%%%%%%%%%%%%%%%%%%%%%%%%%%
\end{document}